\documentclass[sigconf,nonacm]{acmart}

\usepackage{algorithm}

\usepackage{algpseudocode}

\usepackage{subcaption}
\usepackage{multirow}
\usepackage[normalem]{ulem}
\AtBeginDocument{%
  }

\setcopyright{acmlicensed}
\copyrightyear{2018}
\acmYear{2018}
\acmDOI{XXXXXXX.XXXXXXX}

\begin{document}
\pagestyle{plain}

\title{\texorpdfstring{SPICE: \underline{S}peculative \underline{P}refetching with Low-Rank Expert Surrogates and Heterogeneous Orchestration  for MoE \underline{I}nference Ac\underline{c}\underline{e}leration}{SPICE: Speculative Prefetching with Low-Rank Expert Surrogates for MoE Inference Acceleration}}


\author{Yongxiang Lyu}
\affiliation{
  \institution{North Carolina State University}
  \country{United States}
}
\email{ylyu9@ncsu.edu}

\author{Ning Li}
\affiliation{
  \institution{Tianjin University}
  \country{China}
}
\email{ning2022@tju.edu.cn}

\author{Bonian Jia}
\affiliation{
  \institution{New York University}
  \country{United States}
}
\email{bj2536@nyu.edu}

\begin{abstract}
Mixture-of-Experts (MoE) models are increasingly used in LLMs because sparse activation decouples model capacity from compute cost. However, the large expert parameter footprint often exceeds GPU memory capacity, making inference latency dominated by the host-to-device PCIe transfers for expert loading. To address these challenges, this paper presents SPICE, a speculative prefetching framework for MoE offloading that combines lightweight expert prediction with confidence-aware CPU-GPU orchestration. 
On one hand, SPICE builds a lightweight draft model aligned with the target MoE architecture, using a confidence-aware adaptive lookahead algorithm to prefetch high-confidence experts. On the other hand, when speculative predictions miss, SPICE switches to a cost-aware CPU-GPU heterogeneous orchestration: low-confidence misses are approximated by the resident shared expert with low rank expert (LoRE) surrogates, while exact residual work is offloaded to the CPU and executed asynchronously in parallel with ongoing GPU computation. Evaluated on DeepSeek-V2-Lite and Qwen2-57B-A14B across diverse GPU platforms, SPICE achieves up to 3.12$\times$ speedup in Time Per Output Token  (TPOT) with minimal quality loss, showing that effective MoE offloading requires not only predicting future experts, but also deciding which misses deserve approximation, which require exact recovery, and where exact residual work should execute. Our code will be available at
\url{https://anonymous.4open.science/r/SPICE}.
\end{abstract}



\maketitle
\begin{figure}[!t]
  \centering
  \includegraphics[width=\linewidth]{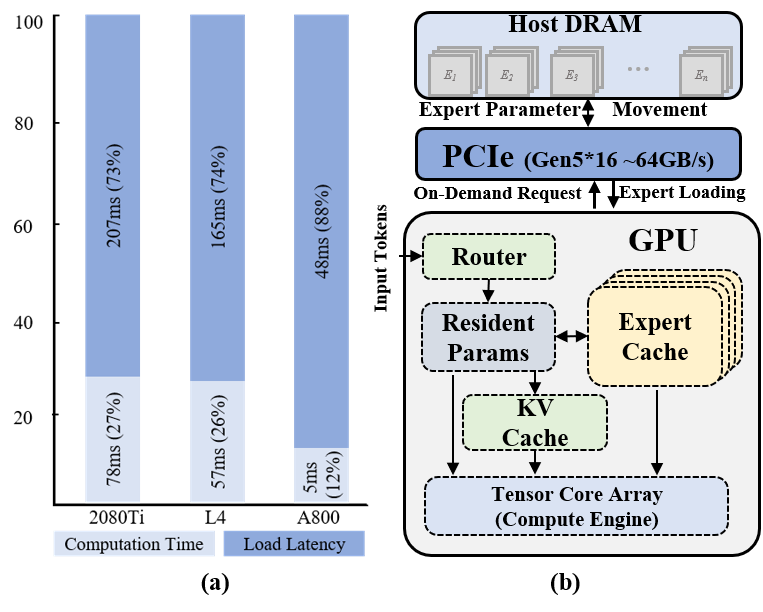}
  \caption{The I/O bottleneck in MoE offloading. (a) Latency breakdown for DeepSeek-V2-Lite\cite{liu2024deepseek} across different GPUs, illustrating that PCIe weight transfer dominates the end-to-end inference time, fundamentally transforming MoE from compute-bound to I/O-bound. (b) The hardware architecture and data movement pipeline of a single MoE layer during offloading.}
  \label{bottleneck}
  \Description{figure1}
\end{figure}

\section{Introduction}
Mixture-of-Experts (MoE) \cite{shazeer2017outrageously, fedus2021switch}
has emerged as a key architecture for scaling large language models (LLMs) by substantially increasing model capacity through sparse activation.
Unlike dense models, MoE activates only a subset of experts per token, allowing the total parameter count to scale without a proportional increase in per-token computation.

Despite the computational efficiency of sparse activation, deploying MoE models within practical GPU memory budgets is constrained by memory capacity.
Fig.~\ref{bottleneck} (a) quantifies one of the bottlenecks: across three representative GPUs, expert loading latency consistently dominates per-layer execution time, accounting for 73--88\% of the total latency, while actual computation occupies only 12--27\%. As illustrated in Fig.~\ref{bottleneck} (b), the non-expert parameters reside permanently in GPU memory alongside the KV cache, while the aggregate size of all expert FFN weights vastly exceeds the high-bandwidth memory (HBM) capacity of a single GPU. This necessitates offloading experts to host DRAM and loading them on demand through the PCIe bus during inference. As a result, MoE inference is fundamentally transformed from a compute-bound to an I/O-bound problem\cite{zhou2025floe}. 

To mitigate this bottleneck, prior work has explored multiple directions. Operator-level and model quantization reduce computational FLOPs and model size\cite{chenmoequant,sheng2023flexgen,qmoe}, yet they do not fundamentally address the memory bandwidth constraint when experts remain off-chip. A natural way to mitigate this stall is expert prefetching: instead of waiting until the router requests an expert, the system predicts likely future experts and transfers their weights from host memory to GPU memory ahead of use \cite{specmd,he2024expertflow,adapmoe}. When the prediction is correct and the transfer completes within the available compute window, prefetching overlaps host-to-device communication with GPU execution and hides part of the I/O latency\cite{popfetcher,li2023accelerating,adapmoe}.
However, prefetching alone is insufficient for low-latency offloaded MoE serving. Its benefit depends on three coupled conditions that are difficult to satisfy simultaneously. First, the prefetcher must predict future expert usage under token- and layer-dependent routing dynamics; the fixed lookahead depths can either miss useful experts or issue stale transfers. Second, when a predicted expert is wrong, recovery falls back to critical-path I/O, requiring the system to reload the missing expert weights before execution can proceed. This recovery cost can erase the latency benefit that prefetching was meant to provide. Third, aggressive lookahead consumes both PCIe bandwidth and the limited GPU-resident expert cache, so speculative transfers may delay urgent demand transfers or evict useful experts.
To address these challenges, we propose SPICE, a novel MoE inference framework designed for inference acceleration. Our contributions are:

\begin{itemize}
\item \textbf{Structurally aligned speculative expert prefetching.} SPICE uses a lightweight draft model aligned with the target MoE to predict future expert accesses, and applies confidence-aware adaptive lookahead to prefetch likely experts.

\item \textbf{Resident low-rank surrogate substitution for low
-confidence misses.}
SPICE uses resident shared experts and Low-Rank Expert (LoRE) surrogates to approximate low-confidence missed experts without stalling on synchronous reloads.

\item \textbf{CPU-GPU heterogeneous orchestration.}
For misses requiring exact recovery, SPICE chooses between loading the expert onto the GPU and executing it on the CPU using host-resident weights.  By overlapping exact CPU execution with asynchronous expert transfers and ongoing GPU computation, SPICE reduces the critical-path latency caused by prefetching misses.
\end{itemize}

\section{Background}
\begin{figure}[t]
  \centering
  \includegraphics[width=\linewidth]{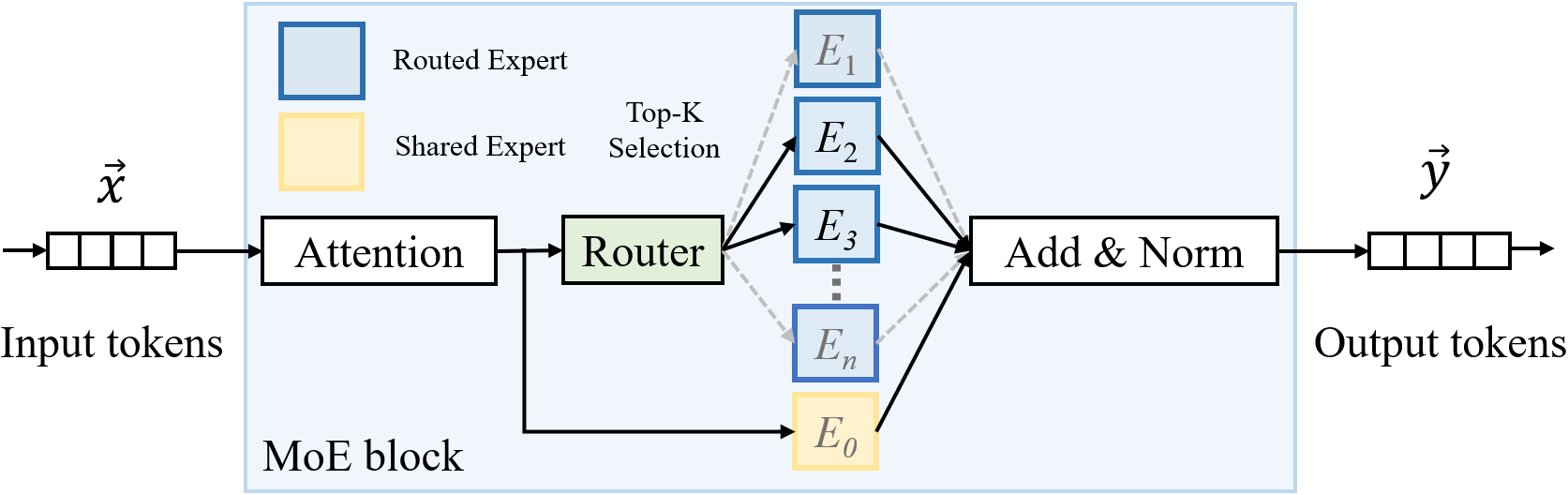}
  \caption{Architecture of a Mixture-of-Experts. A router selects the top-k experts from a pool of routed experts based on the token features. DeepSeek-V2\cite{liu2024deepseek} and Qwen-MoE\cite{qwen_moe} families introduce shared experts in each MoE layer.}
  \label{moe}
  \Description{MoE}
\end{figure}

\subsection{Mixture-of-Experts}
As shown in Fig.~\ref{moe}, an MoE block consists of $n$ expert networks $\{E_1, E_2, \ldots, E_n\}$. The shared experts are always activated, while the router selects the top-K routed experts for each token. The output of all selected experts is then aggregated and weighted.

\begin{figure}[t]
  \centering
  \includegraphics[width=\linewidth]{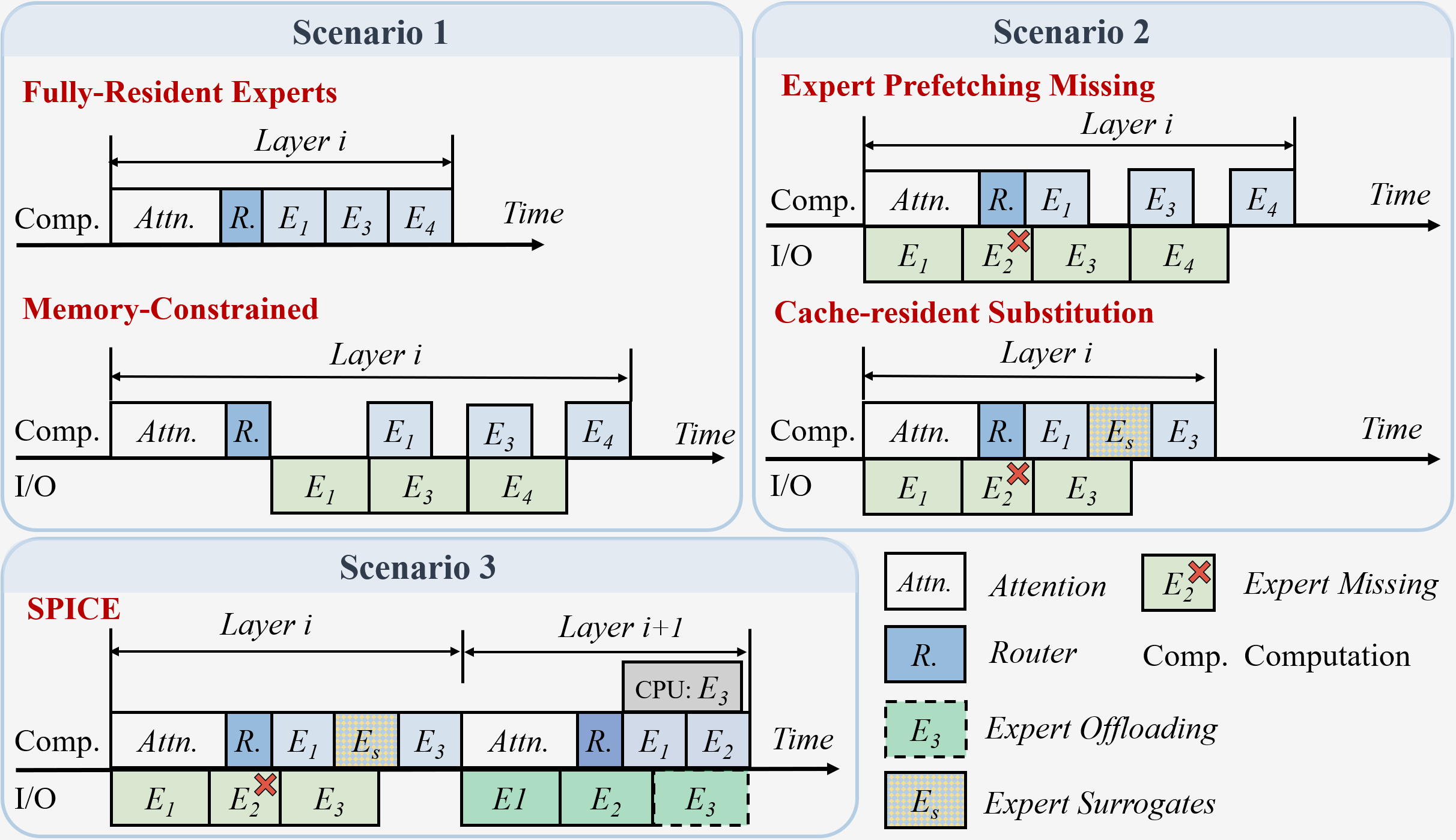}
  \caption{Timeline comparison of memory-constrained MoE execution. Scenario 1 contrasts fully resident execution and on-demand expert
  loading. Scenario 2 shows how a prefetch miss places recovery on the critical path and motivates our LoRE surrogate
  recovery. Scenario 3 illustrates SPICE, where missing experts can be handled through surrogate computation or exact CPU-side
  execution over host-resident experts. }
  \Description{timeline}
\label{timeline}
\end{figure}

\subsection{Expert Prefetching}
Given the hidden state $\mathbf{h}_{t}^{(l)}$, the router selects a sparse expert set:
\begin{equation}
    \mathcal{S}_{t}^{(l)}
    =
    \operatorname{TopK}\!\left(
    \operatorname{Router}^{(l)}(\mathbf{h}_{t}^{(l)}), K
    \right).
\end{equation}
If a selected expert in $\mathcal{S}_{t}^{(l)}$ is not resident in GPU memory, its weights must be transferred from host memory before its FFN can execute. The critical path of a cache miss therefore includes routing, expert-weight transfer, expert computation, and output aggregation.
Expert prefetching attempts to remove weight transfer from this critical path. A predictor estimates the experts required by future layers, and the runtime transfers their weights while earlier layers are still executing. However, prediction alone does not determine a good transfer schedule: unnecessary prefetches consume PCIe bandwidth and cache capacity, while missed predictions still require recovery on the critical path. The runtime must therefore decide not only which experts to prefetch, but also when to transfer them and where to execute missed expert computation.

\subsection{Miss Recovery and CPU--GPU Orchestration}
 When prefetching misses a non-resident expert,
 the runtime must recover its computation without delaying subsequent layers. One option is weight migration, which transfers the expert weights to the GPU and executes the complete expert FFN there. The second is a compute-near-weights path, which transfers the corresponding activation slice to the CPU, evaluates the same expert using host-resident weights, and sends the output back to the GPU for aggregation. Both paths preserve the target model semantics; they differ only in the amount of PCIe traffic they create, the compute device they occupy, and where synchronization delay appears~\cite{CG-MOE}.
Miss recovery therefore requires a joint scheduling decision: the runtime must determine whether to move weights or computation, while coordinating recovery with prefetching and ongoing GPU execution. SPICE treats this decision as a CPU–GPU orchestration problem. It jointly schedules expert prefetching, GPU computation, and asynchronous CPU recovery.

\section{Motivation}
Memory-constrained MoE serving is bottlenecked by expert availability: each layer requires its selected experts before computation, but most expert weights reside in host memory. Fig.~\ref{timeline} illustrates three runtime bottlenecks.

\noindent\textbf{Scenario 1: memory-constrained MoE serving and the need for prefetching.}
 As shown in Figure~\ref{timeline} scenario 1, keeping all experts GPU-resident provides a latency lower bound but is infeasible for large MoE models, particularly as the KV cache grows. Loading missing experts on demand places PCIe transfers directly on the critical path, motivating prefetching that overlaps transfers with earlier GPU computation.

\vspace{1.0ex}
\noindent\textbf{Scenario 2: prefetch misses and the need for cache-resident substitution.} Prefetching hides latency only when predicted experts arrive before use. A miss instead triggers cache eviction and synchronous expert loading. Because shared experts and lightweight surrogates remain GPU-resident, they can provide immediate approximate recovery, trading controlled error for lower blocking latency.


\vspace{1.0ex}
\noindent\textbf{Scenario 3: PCIe Contention Requires Heterogeneous Orchestration.} Aggressive lookahead can saturate PCIe bandwidth, causing even correctly predicted experts to arrive late. Under a limited transfer budget, the runtime must decide how each miss is served. SPICE therefore combines surrogate-based recovery with asynchronous CPU execution over host-resident expert weights, avoiding unnecessary weight transfers while preserving exact execution when required.
\section{SPICE: Methodology and System Design}

The preceding analysis motivates SPICE, a memory-constrained MoE serving system designed to keep decoding progress from being serialized by expert availability. The key idea is to combine speculative expert prefetching with cache-resident substitution and CPU--GPU heterogeneous execution.

\begin{figure}[t]
  \centering
  \includegraphics[width=\linewidth]{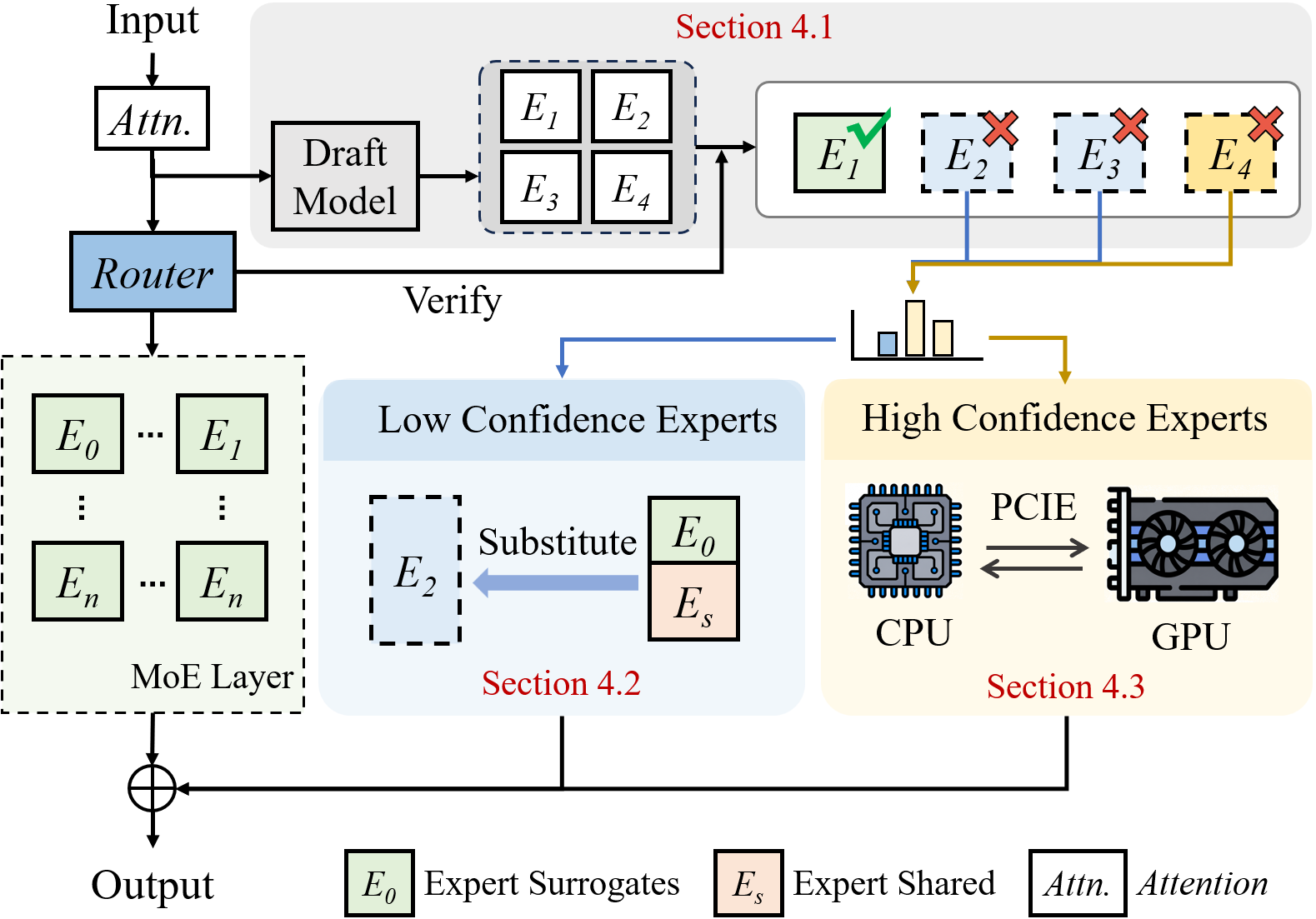}
  \caption{The system overview of SPICE.
}
\label{workflow}
\Description{figure4}
\end{figure}

Figure~\ref{workflow} presents the overall workflow of SPICE.
SPICE first predicts upcoming expert usage and prefetches expert weights to overlap PCIe transfers with GPU computation. When prediction fails, instead of synchronously evicting cached experts and reloading the missing weights on the critical path, SPICE exploits same-layer expert similarity and the shared-expert structure of MoE models to construct cache-resident expert surrogates that approximate or prune the missing expert computation. Meanwhile, selected expert computations are executed on the CPU using host-resident weights, allowing CPU execution and GPU surrogate execution to proceed in parallel. Together, these mechanisms reduce blocking stalls caused by memory limits, prefetch misses, and PCIe contention.

\subsection{Speculative Prefetching Strategy}
Unlike token-level speculative decoding, this mechanism does not predict future output tokens or alter the decoding trajectory; it only predicts the expert sets that are likely to be activated by intra and inter layers.
Specifically, at decoding step \(t\), SPICE anchors the lightweight draft predictor at the target hidden state of the current layer \(l\). It then recursively forecasts expert usage over a bounded lookahead horizon using layer-specific low-rank state transitions. At each prediction depth, the corresponding router combines the predicted hidden state with a compact context accumulated from earlier routing predictions to generate the future Top-\(K\) expert set. These predicted sets trigger speculative prefetches, while the rollout terminates early when prediction confidence falls below a predefined threshold.
Let $\hat{\pi}_{t}^{(l+\delta)} \in [0,1]^N$ denote the predicted routing distribution for future layer $l+\delta$, and let
\begin{equation}
    \hat{\mathcal{S}}_{t}^{(l+\delta)}
    = \mathrm{TopK}(\hat{\pi}_{t}^{(l+\delta)}, K)
\end{equation}
be the corresponding predicted Top-$K$ expert set. SPICE converts these predictions into asynchronous prefetch intents:
\begin{equation}
    \mathcal{Q}_{\mathrm{prefetch}}
    =
    \bigcup_{\delta=1}^{H}
    \hat{\mathcal{S}}_{t}^{(l+\delta)},
\end{equation}
where $H$ is the speculative lookahead horizon. The runtime issues host-to-device copies for experts in $\mathcal{Q}_{\mathrm{prefetch}}$ through a bounded-depth, low-priority copy queue, preventing speculative traffic from monopolizing the PCIe copy engine.

The remaining question is how far the system should look ahead. A fixed prefetch horizon is poorly matched to MoE decoding: a shallow horizon cannot hide long PCIe transfers, while an overly deep horizon may pollute the limited GPU cache with stale expert predictions when routing uncertainty increases. SPICE therefore adapts the prefetch depth using two signals: the hardware overlap requirement and the predictor confidence.

For a prefetch count $P$ per layer, the transfer time is $P \cdot M_e / B_{\mathrm{PCIe}}$, where $M_e$ is the memory footprint of one expert and $B_{\mathrm{PCIe}}$ is the measured host-to-device bandwidth. Given the per-layer compute window
$T_{\mathrm{comp}} = T_{\mathrm{attn}} + K \cdot T_{\mathrm{expert}}$, the minimum lookahead depth required to hide this transfer is
\begin{equation}
    H_{\min}
    =
    \left\lceil
    \frac{P \cdot M_e}
    {B_{\mathrm{PCIe}} \cdot T_{\mathrm{comp}}}
    \right\rceil
    =
    \left\lceil
    \frac{P \cdot M_e}
    {B_{\mathrm{PCIe}} \cdot (T_{\mathrm{attn}} + K \cdot T_{\mathrm{expert}})}
    \right\rceil .
    \label{eq:h_min}
\end{equation}
This bound specifies how early a reliable prefetch should be issued so that expert migration can proceed as background work instead of becoming a demand-driven stall.

\begin{figure}[t]
  \centering
  \includegraphics[width=0.95\linewidth]{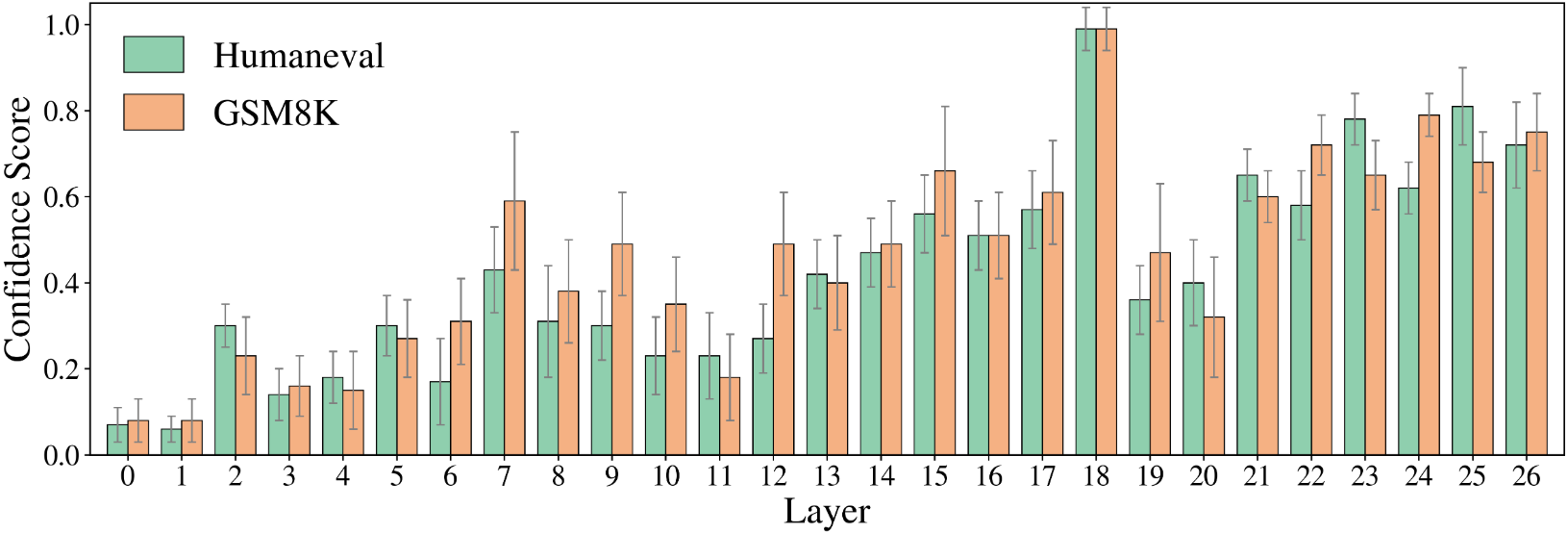}
  \caption{Routing confidence across layers on HumanEval and GSM8K. Each bar represents the mean confidence $\mathcal{C}(\hat{\pi}^{(l)})$ (Eq.~\ref{eq:confidence}), with error bars indicating one standard deviation. Confidence varies substantially across layers and workloads, motivating depth-adaptive halting. $\tau$ denotes the threshold used in Algorithm~\ref{alg:adaptive_prefetch}.}
  \label{fig:confidence}
  \Description{figure5}
\end{figure}

SPICE then prevents the horizon from growing beyond what the prediction quality can justify. Let $\hat{\pi}_{t}^{(l)} \in [0,1]^N$ denote the draft model's predicted routing distribution for layer $l$ at decoding step $t$, and let $\hat{\mathcal{S}}_{t}^{(l)}$ be the corresponding Top-$K$ expert set. We define the prediction confidence as the probability mass assigned to the selected experts:
\begin{equation}
    \mathcal{C}(\hat{\pi}_{t}^{(l)}) =
    \sum_{i \in \hat{\mathcal{S}}_{t}^{(l)}} \hat{\pi}_{t,i}^{(l)} .
    \label{eq:confidence}
\end{equation}As shown in Figure~\ref{fig:confidence}, prediction confidence varies substantially across layers and workloads, motivating an adaptive rather than fixed lookahead depth. After reaching $H_{\min}$, SPICE stops further expansion when $C(\hat{\pi}_t^{(l)}) < \tau$; otherwise, it continues up to $H_{\max}$. This limits uncertain prefetches without changing the target routing policy. Misses are handled by the recovery path described in Section~4.2. Algorithm~\ref{alg:adaptive_prefetch} summarizes this procedure. SPICE combines these two signals into an adaptive halting mechanism (Algorithm~\ref{alg:adaptive_prefetch}). The algorithm operates as follows:


\begin{algorithm}[ht]
\caption{Speculative Prefetching in SPICE}
\label{alg:adaptive_prefetch}
\begin{algorithmic}[1]
\State \textbf{Input:} Current hidden state $x^{(\mathrm{current})}$, target-model attention and routers, LoRE surrogates
\State \textbf{Parameters:} Minimum lookahead depth $l_{\min}$, maximum lookahead depth $l_{\max}$, confidence threshold $\tau$
\State \textbf{Output:} Expert prefetch queue $\mathcal{Q}_{\mathrm{prefetch}}$
\Statex
\State Initialize $l \leftarrow 1$ and $\hat{h}^{(0)} \leftarrow x^{(\mathrm{current})}$
\State $\mathcal{Q}_{\mathrm{prefetch}} \leftarrow \emptyset$

\While{$l \leq l_{\max}$}
    \State \Comment{Step 1: Predict expert demand}
    \State $\hat{\pi}^{(l)} \leftarrow
        \operatorname{router}^{(l)}(\hat{h}^{(l-1)})$
    \State $\hat{\mathcal{S}}^{(l)} \leftarrow
        \operatorname{TopK}(\hat{\pi}^{(l)}, K)$

    \State \Comment{Step 2: Enqueue the current prediction}
    \State $\mathcal{Q}_{\mathrm{prefetch}}.
        \operatorname{enqueue}(\hat{\mathcal{S}}^{(l)})$

    \State \Comment{Step 3: Determine whether to continue speculation}
    \State $\mathcal{C}^{(l)} \leftarrow
        \sum_{i \in \hat{\mathcal{S}}^{(l)}} \hat{\pi}^{(l)}_i$
    \If{$l \geq l_{\min}$ \textbf{and} $\mathcal{C}^{(l)} < \tau$}
        \State \textbf{break}
        \Comment{Retain the current prefetch and stop deeper speculation}
    \EndIf

    \State \Comment{Step 4: Project the draft state to the next depth}
    \State $\hat{h}^{(l)} \leftarrow
        \operatorname{Attn}^{(l)}(\hat{h}^{(l-1)})
        + \displaystyle\sum_{i \in \hat{\mathcal{S}}^{(l)}}
        \hat{\pi}^{(l)}_i
        \tilde{E}^{(l)}_i(\hat{h}^{(l-1)})$
    \State $l \leftarrow l + 1$
\EndWhile

\State \textbf{Issue asynchronous PCIe transfers for experts in }
    $\mathcal{Q}_{\mathrm{prefetch}}$
\State \textbf{Trigger the target-model execution pipeline}
\end{algorithmic}
\end{algorithm}

\subsection{Low-Confidence Substitution with Shared Expert and LoRE Surrogates}
The second component of SPICE handles low-confidence misses through approximate substitution rather than exact expert execution. We focus on target MoE architectures that provide a native shared expert in each MoE layer. In such models, the shared expert is always available on the GPU and provides a stable layer-wise common transformation. SPICE uses this shared expert as the base computation and augments it with a lightweight low-rank expert surrogate, called LoRE, to approximate the missing routed expert when the routed expert has low confidence.

For layer $l$, let $S^{(l)}(\cdot)$ denote the native shared expert and $E_e^{(l)}(\cdot)$ denote routed expert $e$. Given the token hidden state $h^{(l)}$, the exact routed expert output is
  \begin{equation}
  y_e^{(l)} = E_e^{(l)}(h^{(l)}).
  \end{equation}
  Instead of reconstructing the full expert, SPICE learns a low-rank residual correction on top of the shared expert:
  \begin{equation}
  \tilde{y}_e^{(l)}
  =
  S^{(l)}(h^{(l)})
  +
  \Delta_e^{(l)}(h^{(l)}),
  \end{equation}
  where $\Delta_e^{(l)}(\cdot)$ is the LoRE surrogate for expert $e$ at layer $l$.

  The LoRE surrogate is obtained offline from target-model traces. During calibration, we run the target MoE on a held-
  out set and record the hidden states, router decisions, shared expert outputs, and exact routed expert outputs. For
  each layer-expert pair $(l,e)$, SPICE fits a low-rank residual module to approximate the difference between the routed
  expert and the shared expert:
  \begin{equation}
  \Delta_e^{(l)}(h)
  \approx
  E_e^{(l)}(h) - S^{(l)}(h).
  \end{equation}
  In our implementation, this residual module is parameterized as a low-rank transformation,
  \begin{equation}
  \Delta_e^{(l)}(h) = B_e^{(l)} A_e^{(l)} h,
  \end{equation}
  where $A_e^{(l)} \in \mathbb{R}^{r \times d}$ and $B_e^{(l)} \in \mathbb{R}^{d \times r}$ with $r \ll d$. The
  parameters are trained by minimizing the residual approximation error over calibration tokens routed to expert $e$:
  \begin{equation}
  \min_{A_e^{(l)},B_e^{(l)}}
  \sum_{h \in \mathcal{D}_{l,e}}
  \left\|
  E_e^{(l)}(h)
  -
  S^{(l)}(h)
  -
  B_e^{(l)} A_e^{(l)} h
  \right\|_2^2 .
  \end{equation}
  This construction makes LoRE a correction term rather than a standalone expert. The shared expert captures the common
  feed-forward transformation of the layer, while LoRE captures the expert-specific deviation using only a small number
  of parameters.

  At runtime, this substitution path is used only for low-confidence missing experts. SPICE first obtains the target
  router's selected experts and their gate scores. If a selected expert is already resident on the GPU or has been
  prefetched, SPICE executes it exactly. If the expert is missing, SPICE checks its routing confidence. Low-confidence
  misses are substituted by the shared-expert-plus-LoRE output and therefore do not trigger either a host-to-device
  expert transfer or CPU expert execution:
  \begin{equation}
  \tilde{y}_e^{(l)}
  =
  S^{(l)}(h^{(l)})
  +
  B_e^{(l)} A_e^{(l)} h^{(l)} .
  \end{equation}
SPICE handles each prefetch miss according to its estimated contribution. Low-confidence misses use a resident LoRE surrogate, avoiding synchronous expert loading at a controlled accuracy cost. High-contribution misses require exact recovery: the runtime either loads the expert onto the GPU or executes it on the CPU using host-resident weights. These paths respectively reduce blocking transfers and preserve exact computation when approximation is unsuitable.

\subsection{CPU-GPU Heterogeneous Orchestration}

After speculative prefetching and low-confidence substitution, a residual bottleneck remains: some router-selected experts are still absent from the GPU but cannot be safely approximated. SPICE therefore handles these remaining misses with a CPU--GPU heterogeneous scheduler, which decides whether each expert should be fetched to the GPU or executed exactly on the CPU using host-resident weights.

Let $\mathcal{M}^{(l)}$ denote the set of non-substituted missed experts at layer $l$. SPICE partitions this set into two disjoint subsets:
\begin{equation}
    \mathcal{M}^{(l)}
    =
    \mathcal{M}_{\mathrm{fetch}}^{(l)}
    \cup
    \mathcal{M}_{\mathrm{cpu}}^{(l)},
    \qquad
    \mathcal{M}_{\mathrm{fetch}}^{(l)}
    \cap
    \mathcal{M}_{\mathrm{cpu}}^{(l)}
    =
    \emptyset .
\end{equation}
Experts in $\mathcal{M}_{\mathrm{fetch}}^{(l)}$ are transferred over H2D and executed on the GPU once the copy completes, while experts in $\mathcal{M}_{\mathrm{cpu}}^{(l)}$ are executed on the CPU using host-resident weights. The CPU path is exact: it uses the same target expert weights and does not invoke surrogate experts, alter routing decisions, or change the target model computation graph.
\begin{figure}[t]
  \centering
  \includegraphics[width=\linewidth]{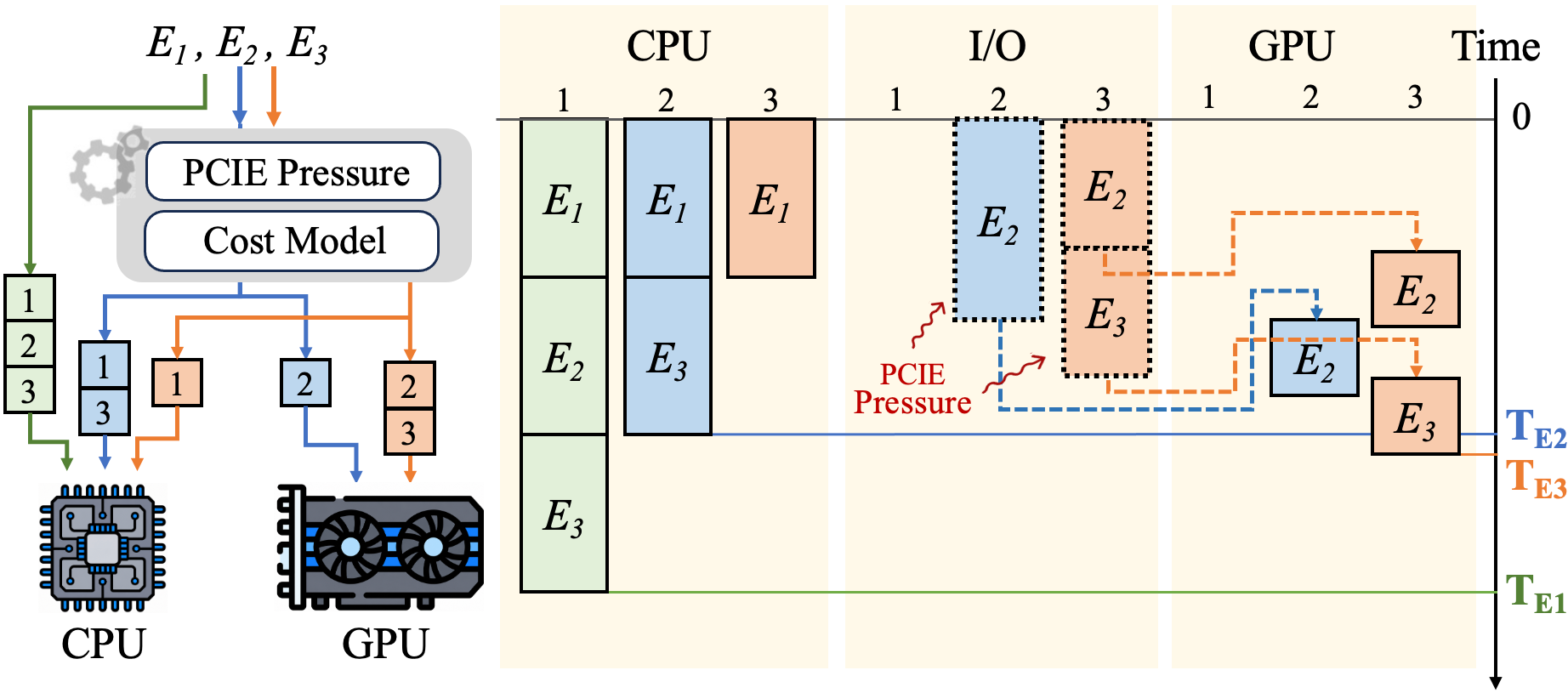}
  \caption{CPU--GPU heterogeneous orchestration under PCIe pressure. The cost model maps events $E_1$, $E_2$, and $E_3$ to CPU and GPU resources, while dashed paths expose PCIe transfers and contention that delay $E_2$ and $E_3$.
}
\label{cpu-gpu}
\Description{figure6}
\end{figure}
The mixed scheduler determines this split using the current PCIe pressure and a per-layer cost table. SPICE assigns a missed expert to GPU fetching only when the residual transfer is expected to reduce end-to-end recovery time despite ongoing prefetch traffic. Otherwise, the expert is assigned to CPU residual execution. In implementation, the scheduler compares all-CPU service with candidate split plans and selects a GPU-fetch count only when the estimated split cost improves over all-CPU execution by a safety margin. The selected experts are fetched through a high-priority residual stream, while the remaining experts are executed on the CPU. As shown in Fig.~\ref{cpu-gpu}, SPICE dynamically assigns missed experts to either GPU fetch-and-compute or exact CPU residual execution. Execution proceeds in three steps. First, SPICE stages the current hidden state in pinned host memory, launches CPU residual execution, and issues asynchronous H2D transfers for the selected GPU-fetch experts. Second, CPU expert computation progresses in parallel with the copy engine, converting part of the residual miss cost into useful host-side work. Third, after GPU transfers complete, SPICE executes the resident and fetched experts on the GPU and merges their outputs with the CPU expert outputs to form the next-layer hidden state. This organization hides part of the copy latency behind exact CPU computation while preserving target-model semantics for all non-substituted misses. This design separates approximate and exact recovery. 
By contrast, CPU residual execution preserves exact computation for non-substituted misses when GPU-side weight transfer would be too costly.
\section{Experiments and Results}
\subsection{Experimental Setup}
\label{sec:setup}

\textbf{Hardware and Models.} We evaluate SPICE under three GPU configurations: (a) an NVIDIA 5090 with PCIe 4.0 x8 and 128\,GB DDR5 host DRAM, (b) an NVIDIA RTX 4060 with PCIe 4.0 x8 and 16\,GB DDR4 host DRAM and (c)NVIDIA A800 80GB PCIe GPUs with PCIe 4.0 x16 links and about 512 GB DDR4 host DRAM on an Intel Xeon Gold 5320 platform. We evaluate on two MoE models: DeepSeek-V2-Lite~\cite{liu2024deepseek} and Qwen2-57B-A14B~\cite{qwen_moe}.

\textbf{Baselines.} We compare SPICE with several representative MoE inference baselines. MoE-On-Demand (Naive): all experts are stored on the CPU. Experts are dynamically loaded into the GPU when selected for inference. AdapMoE\cite{adapmoe} is a GPU-centric scheduling framework that reduces on-demand loading overhead via adaptive
expert prefetching and caching. CG-MoE\cite{CG-MOE} utilizes both CPU and GPU through joint expert scheduling and cache hierarchy management.

\textbf{Datasets.}
We evaluate inference quality on four benchmarks: MT-Bench~\cite{mt-bench}, LongBench~\cite{bai2024longbench}, HumanEval~\cite{human}, and GSM8K~\cite{cobbe2021training}.


\textbf{Measurement protocol.} We measure energy during end-to-end MoE inference, including model computation, expert transfers, draft-model inference, and CPU-side expert recovery. GPU energy is obtained by integrating NVML power samples collected at 10 Hz, while CPU-package and DRAM energy are measured using Intel RAPL counters across all CPU sockets over the same decoding interval. We define measured platform energy as the sum of GPU, CPU-package, and DRAM energy, and report energy per output token by dividing this sum by the number of generated tokens.

\subsection{End-to-End Inference Performance}
\begin{figure}[!t]
  \centering
    \includegraphics[width=\linewidth]{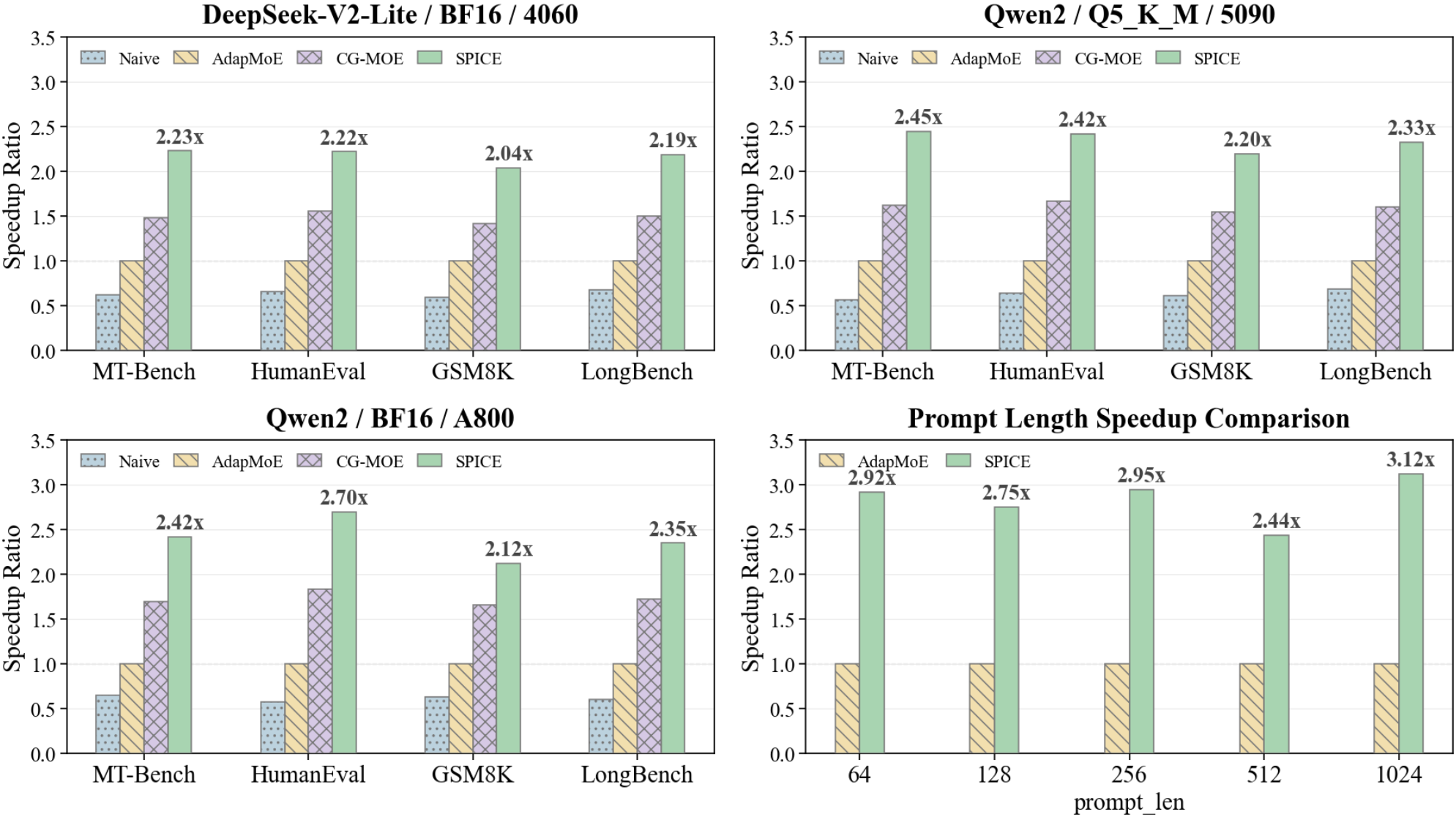}
  \caption{End-to-end TPOT speedup relative to AdapMoE across workloads, model--GPU configurations. Speedup is
computed as $\mathrm{TPOT}_{\mathrm{method}} / \mathrm{TPOT}_{\mathrm{AdapMoE}} $; Higher is better.}
    \label{fig:tpot}
  \Description{7}
\end{figure}

We compare the end-to-end inference latency of SPICE against baselines across diverse models and configurations, shown in the Fig~\ref{fig:tpot}.
SPICE achieves 2.04–2.70$\times$ speedup over AdapMoE and consistently outperforms CG-MoE. Its speedup remains between 2.44× and 3.12× as the prompt length increases from 64 to 1,024 tokens, demonstrating robust gains across the evaluated context lengths.
Across two MoE models in Table~\ref{tab:accuracy}, SPICE incurs only a 3.0–3.5 percentage-point accuracy drop on GSM8K and a 2.3–3.9 point Pass@1 drop on HumanEval. These results indicate that SPICE’s approximation introduces an acceptable quality trade-off for its inference-efficiency gains.
\begin{table}[!t]
\centering
\caption{Inference quality on GSM8K and HumanEval under different model precisions. $^\dagger$BF16; $^\ddagger$Q5\_K\_M.}
\label{tab:accuracy}
\setlength{\tabcolsep}{4pt}
\small
\begin{tabular}{lcccccc}
\toprule
\multirow{2}{*}{Model} & \multicolumn{3}{c}{GSM8K (Acc.)} & \multicolumn{3}{c}{HumanEval (Pass@1)} \\
\cmidrule(lr){2-4} \cmidrule(lr){5-7}
 & Naive & SPICE & $\Delta$ & Naive & SPICE & $\Delta$ \\
\midrule
DeepSeek-V2-Lite$^\dagger$           & 70.8 & 67.5 & $-3.3$ & 55.1 & 52.7 & $-2.4$ \\
Qwen2-57B-A14B$^\dagger$   & 87.6 & 84.1 & $-3.5$ & 74.5 & 70.6 & $-3.9$ \\
Qwen2-57B-A14B$^\ddagger$  & 83.2 & 80.2 & $-3.0$ & 73.2 & 70.9 & $-2.3$ \\
\bottomrule
\end{tabular}
\end{table}

\subsection{Energy and PCIe Traffic Analysis}
\label{sec:energy}

\textbf{Energy analysis.} Utilization is measured over the complete decoding interval and normalized to the measured peak H2D bandwidth of each GPU. 
Table~\ref{tab:energy} shows that lower power does not necessarily imply lower energy. MoE-On-Demand (Naive) draws the least power but consumes the most energy comsumption at 5.51 J/token because synchronous loading prolongs execution. The other Policies that reduce energy by 31–41\% despite drawing more power. SPICE achieves 3.48 J/token while reducing fallback to 3.0\% and H2D traffic to 54.7 GB.

\textbf{H2D traffic and fallback.} Table~\ref{tab:energy} reveals a trade-off between transfer volume and expert availability. LRU and CG-MoE minimize H2D traffic at 30.2 GB, but both leave 15.7\% of expert accesses to fallback. AdapMoE reduces fallback to 6.1\% by prefetching more aggressively, increasing traffic to 75.7 GB. SPICE achieves the lowest fallback rate of 3.0\% with 54.7 GB of traffic, reducing H2D traffic by 28\% and fallback by 51\% relative to AdapMoE. Thus, SPICE improves expert availability without incurring proportional transfer overhead.

\begin{figure}[h]
  \centering
  \includegraphics[width=\linewidth]{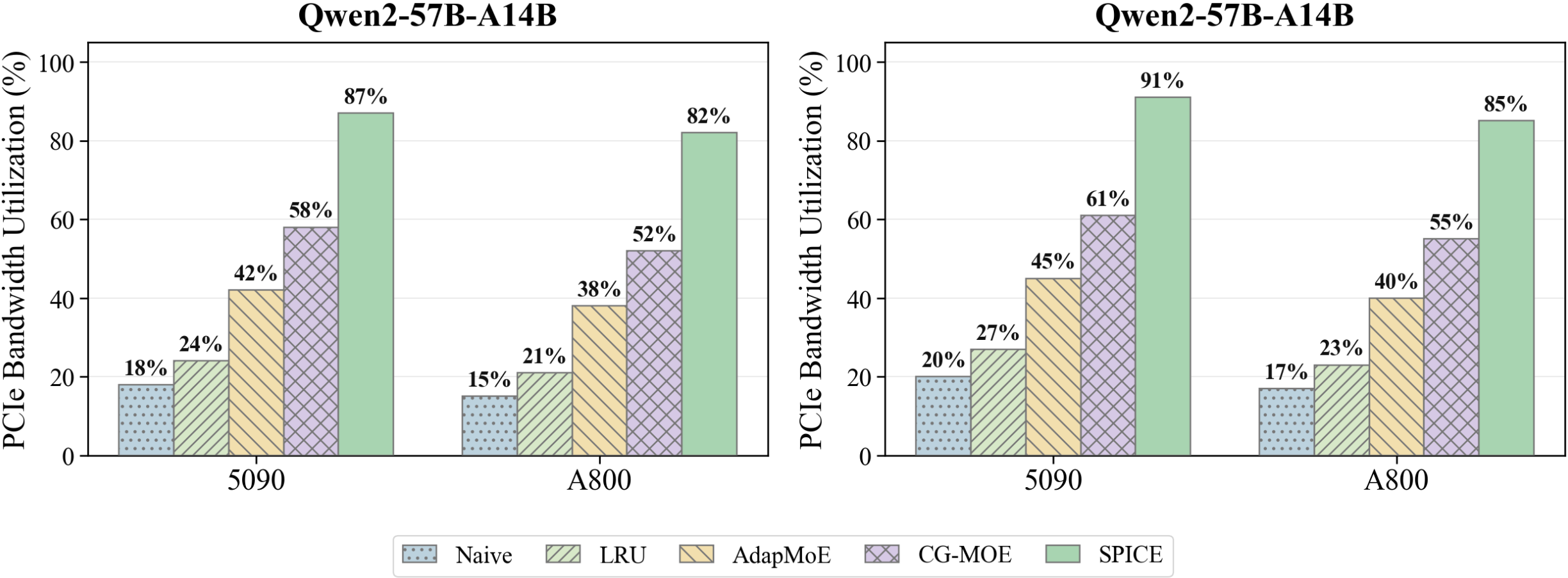}
  \caption{Average PCIe H2D bandwidth utilization during end-to-end Qwen2-57B-A14B inference under (a) BF16 and (b) Q5\_K\_M precision.}
  \label{fig:bandwidth}
  \Description{figure8}
\end{figure}

\textbf{PCIe utilization} Figure~\ref{fig:bandwidth} shows that SPICE consistently makes greater use of the available H2D bandwidth across both model precisions and GPU platforms. Its utilization ranges from 82\% to 91\%, compared with 15–20\% for Naive, 21–27\% for LRU, 38–45\% for AdapMoE, and 52–61\% for CG-MoE. Together with SPICE’s lower fallback rate, these results are consistent with its multi-layer prefetcher exposing expert transfers early enough to reduce idle PCIe intervals and overlap communication with computation, rather than merely transferring more data.

\begin{table}[t]
  \centering
  \caption{End-to-end energy and H2D traffic for 256 decoding steps. The evaluated configuration uses Top-$K=8$ routing and allocates a 512-slot expert cache. Best and
  second-best values are shown in bold and underlined, respectively.}
  \label{tab:energy}
  \small
  \begin{tabular}{lrrrrr}
    \toprule
    Policy & Fallback & H2D (GB) & Power (W)
           & J/token & Norm. \\
    \midrule
    Naive
      & 100\% & 192.0 & 268.6 & 5.51 & 1.00 \\
    LRU
      & 15.7\% & \textbf{30.2} & 418.1
      & \textbf{3.25} & \textbf{0.59} \\
    CG-MoE
      & 15.7\% & \textbf{30.2} & 558.9
      & 3.53 & 0.64 \\
    AdapMoE
      & \underline{6.1\%} & 75.7 & 587.0
      & 3.78 & 0.69 \\
    SPICE
      & \textbf{3.0\%} & \underline{54.7} & 573.0
      & \underline{3.48} & \underline{0.63} \\
    \bottomrule
  \end{tabular}
\end{table}

\subsection{Ablation Study}
\begin{figure}[h]
  \centering
  \includegraphics[width=\linewidth]{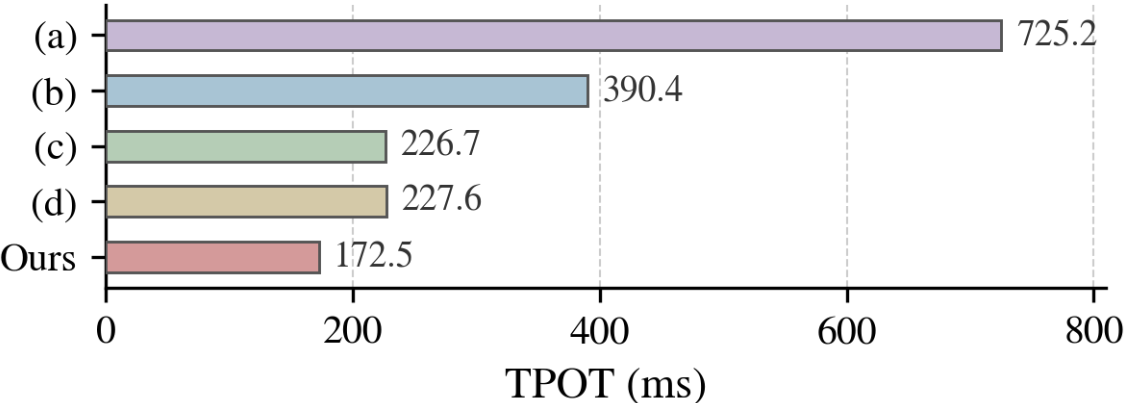}
  \caption{Ablation of SPICE components under the exact-residual setting. We compare four partial variants: (a) deep prediction with fetch-all prefetching; (b) deep prediction with CPU fallback; (c) shallow next-layer prediction with CPU fallback; and (d) shallow next-layer prediction with transfer scheduling.}
\label{abltion}
\Description{figure9}
\end{figure}

Figure~\ref{abltion} reports TPOT for four ablated SPICE variants under the exact-residual setting. The full system achieves 172.5 ms, a 23.9\% reduction over the best ablation, showing that cross-layer prediction and confidence-aware transfer scheduling are complementary.

\section{Conclusion}
In this work, we investigate the challenges of MoE inference under memory constraints. We analyze representative deployment scenarios, discuss the corresponding optimization techniques, and identify their limitations. Building on these insights, we present SPICE, which provides a system-level perspective on efficient MoE inference in resource-constrained environments and sets the stage for future work on efficient MoE deployment.


\clearpage

\bibliographystyle{ACM-Reference-Format}
\bibliography{software.bib}
\end{document}